\documentclass[useAMS,usenatbib]{mn2e}
\usepackage{graphicx}
\usepackage{subfigure}
\usepackage{natbib}

\usepackage{epstopdf}
\usepackage{multirow}
\usepackage{amssymb}

\newcommand{\mass}[1]{\langle #1 \rangle_{\rm M}}
\newcommand{\vol}[1]{\langle #1 \rangle_{\rm V}}
\newcommand{\mH}{{\rm H}}
\newcommand{\mHt}{{\rm H_{2}}}
\newcommand{\me}{{\rm e^{-}}}
\newcommand{\Hp}{{\rm H^{+}}}

\title[]{Modeling H$_2$ formation in the turbulent ISM: Solenoidal versus compressive turbulent forcing}
\author[Micic et al.]{Milica Micic$^{1}$\thanks{E-mail:
milica@uni-hd.de}, Simon C. O. Glover$^{1}$, Christoph Federrath$^{1,2,3}$ and Ralf S. Klessen$^{1}$\\
$^{1}$Zentrum f\"{u}r Astronomie der Universit\"{a}t Heidelberg, Institut f\"{u}r Theoretische Astrophysik, Albert-Ueberle-Str. 2, 69120 Heidelberg\\
$^{2}$Ecole Normale Sup\'{e}rieure de Lyon, CRAL, 69364 Lyon, France\\
$^{3}$Monach Centre for Astrophysics (MoCA), School of Mathematical Sciences, Monach University, Vic 3800, Australia}

\begin{document}

\date{\today}

\pagerange{\pageref{firstpage}--\pageref{lastpage}} \pubyear{2011}

\maketitle

\label{firstpage}

\begin{abstract}

We present results from high-resolution three-dimensional simulations of the turbulent interstellar medium that study the influence of the nature of the turbulence on the formation of molecular hydrogen. We have examined both solenoidal (divergence-free) and compressive (curl-free) turbulent driving, and show that compressive driving leads to faster H$_{2}$ formation, owing to the higher peak densities produced in the gas. The difference in the H$_{2}$ formation rate can be as much as an order of magnitude at early times, but declines at later times as the highest density regions become fully molecular and stop contributing to the total H$_{2}$ formation rate. We have also used our results to test a simple prescription suggested by Gnedin et al.~(2009) for modeling the influence of unresolved density fluctuations on the H$_{2}$ formation rate in large-scale simulations of the ISM. We find that this approach works well when the H$_{2}$ fraction is small, but breaks down once the highest density gas becomes fully molecular.

\end{abstract}


\section{Introduction}

All observed Galactic star formation takes place within dense, massive clouds of molecular gas known as giant molecular clouds (GMCs). Understanding how these clouds form and evolve is therefore a crucial part of the study of star formation on galactic scales. In the past, molecular clouds have been seen as quasi-static objects that form stars slowly over a long lifetime where the dynamical evolution of a cloud and the chemical evolution of the gas within it were only loosely coupled and were modeled separately. However, observations provide velocity dispersions documenting the existence of supersonic random motions on scales larger than $\sim$0.1 pc \citep[e.g.][]{gk74,ze74,lar81,my83,per86,sol87,fal92,oml02,hb04,rd11,sch11}. These motions have been associated with compressible turbulence in the interstellar medium (ISM) leading to appreciation that GMCs are highly inhomogeneous and that their formation and evolution are dominated by the effects of supersonic turbulent motions \citep[]{bphv99,rk04,es04,se04}. The dynamical evolution of the clouds is rapid, with a timescale comparable to those of the most important chemical processes such as the conversion of atomic to molecular hydrogen or the freeze-out of molecules onto the surfaces of interstellar dust grains. In this picture, the dynamics and chemistry of the gas are strongly coupled, with one directly influencing the evolution of the other, meaning that they must be modeled together.

The main chemical constituent of the molecular gas is molecular hydrogen, H$_2$, with other molecules such as CO being present only in small amounts, so in practice the study of the formation of molecular gas is usually simply the study of the formation of H$_2$. The molecule forms in the interstellar medium primarily on the surface of dust grains. Its formation in the gas phase by radiative association is highly forbidden due to the molecule's lack of a permanent dipole moment and occurs at a negligibly slow rate \citep[]{gs63}, while the gas phase formation via intermediate molecular ions such as H$^-$ or H$^{+}_{2}$ is strongly suppressed by the interstellar radiation field \citep[]{gl03} and cannot produce molecular fractions much higher than $x_{\rmn{H_{2}}}\simeq10^{-3}$.

Given the relatively slow rate at which H$_2$ forms, it is natural to ask whether it is possible to produce large amounts of H$_2$ quickly enough for a model involving rapid cloud formation to be viable. \citet{gml07b} have shown that dynamical processes such as supersonic turbulence have a great impact on the effective H$_2$ formation rate. The presence of turbulence dramatically reduces the time required to form large quantities of H$_2$. The density compressions created by  supersonic turbulence allow H$_2$ to form rapidly, with large molecular fractions being produced after only 1--2~Myr, consistent with the timescale required by rapid cloud formation models. It is found that much of the H$_2$ is formed in high density gas and then transported to lower densities by the action of the turbulence \citep[]{fed08a}, a phenomenon that certainly has a significant impact on the chemistry of the ISM.

One issue not addressed in the study by \citet{gml07b} was the sensitivity of these results to the nature of the turbulent velocity field. Most of the work that has been done to date on the numerical modeling of molecular cloud turbulence has focussed on either purely solenoidal (i.e.\ divergence-free) turbulence, or weakly compressive turbulence where the solenoidal modes dominate over the compressive (curl-free) modes \citep[see e.g.][]{khm00,kls01,osg01,ls08}. The study by \citet{gml07b} is no exception, as it used the same setup for generating weakly compressive turbulence as in earlier work by \citet{mkbs98} and \citet{ml99}. Recently, however, Federrath and collaborators have performed a number of studies of fully compressive turbulence \citep{fed08b,fed09,schm09,fed10}. They show that compressive turbulence produces a significantly broader spread of densities than solenoidal turbulence with the standard deviation of the density probability distribution functions (PDFs) differing by a factor of 3 at the same rms Mach number and argue that while solenoidal forcing of turbulence is likely to occur in quiescent regions with low star formation rates like in the Polaris Flare and or Maddalena's Cloud, regions with a higher star formation activity are more compatible with compressive turbulence \citep[see also,][]{fed10,br10,pfb11}.

The influence of the wide spread of densities produced by compressively driven turbulence on the rate at which molecular hydrogen forms in the ISM has not previously been investigated, but given the strong density dependence of the H$_{2}$ formation rate, it is plausible that the effect could be large. To address this issue, we have carried out a numerical investigation of the rate at which H$_{2}$ forms in interstellar gas dominated by compressive turbulence, and how this compares to the H$_{2}$ formation rate in gas dominated by solenoidal turbulence. 

The outline of our paper is as follows.  In \S2 we describe our numerical method, paying particular attention to the treatment of chemistry and cooling, as well as the method used to generate and maintain turbulence in the gas. In \S3 we present our results for the H$_{2}$ formation rate, and discuss the distributions of density, temperature and H$_{2}$ abundance generated in the simulations. We also use our results to test the sub-grid scale model for H$_{2}$ formation in turbulent gas put forward by \citet{gtk09}. We close with a summary of our findings in section 4.

\section{Simulations}

\subsection{Numerical method}

\subsubsection{Chemistry and cooling}
\label{chemcool}
Modeling the thermal evolution of the gas in a meaningful fashion and having a full chemical model of the ISM can easily require one to track several hundred different atomic and molecular species involved in several thousand different reactions, even if reactions on grain surfaces are neglected \citep[see e.g. the UMIST Database for Astrochemistry, as described in][]{umist}. This is impractical to include in a 3D hydrodynamical code, since it would have an extreme impact on the code's performance. In order to run time-dependent chemical networks efficiently alongside the dynamical evolution of the system one needs to select a number of chemical species and mutual reactions such that the chemical network can be solved in a short enough time while still adequately describing the overall evolution of the system \citep[see][]{gml07a,gml07b}. For our purposes we need to be able to follow the formation and destruction of H$_{2}$ with a reasonable degree of accuracy. 

We have modified the FLASH v2.5 adaptive mesh refinement code \citep[]{fr00, ca02} to include a detailed atomic/molecular cooling function and a simplified but accurate treatment of the most important hydrogen chemistry. FLASH is a massively parallel code, developed by the Center for Astrophysical Thermonuclear Flashes at the University of Chicago. It has support for a variety of different physical processes, including magneto-hydrodynamical (MHD) flows and self-gravity. FLASH uses the PARAMESH library to manage a block-structured adaptive grid and the Message-Passing Interface (MPI) for parallelization. 

Our modifications add a limited treatment of non-equilibrium chemistry treated in an operator-split fashion \citep[]{gml07a, gml07b}. During each hydro step, the coupled set of chemical rate equations for the fluid are solved using the implicit integrator DVODE \citep[]{dvode}, together with the portion of the internal energy equation dealing with compressional and radiative heating and cooling, under the assumption that the other hydrodynamical variables (e.g. density) remain fixed. The advection of the gas energy density is handled as in the unmodified FLASH code. Chemical abundances are tracked using FLASH's standard tracer field implementation, and consistent multifluid advection \citep{pm99} is used to reduce the advection errors.

By default, the internal energy in FLASH is computed by subtracting the specific kinetic energy from the total specific energy, using the equation
\begin{equation}
\varepsilon = E-\frac{|\rmn{\mathbf{v}}|^{2}}{2},
\label{eq:spinten} 
\end{equation}
where $\varepsilon$ is the specific internal energy, $E$ is the specific total energy and $\mathbf{v}$ is the velocity. In regions where the kinetic energy greatly dominates the total energy due to truncation error this approach can lead to unphysical (e.g.\ negative) internal energies, giving inaccurate values for pressures and temperatures. This problem can be avoided by evolving the internal energy separately, using the equation
\begin{equation}
\frac{\partial\rho\varepsilon}{\partial t}+\nabla\cdot[(\rho\varepsilon+P)\rmn{\mathbf{v}}]-\rmn{\mathbf{v}}\cdot\nabla P=0,
\label{eq:inteng}
\end{equation}
where $\rho$ is the density and $P$ is the gas pressure. The method used within the FLASH code is determined via the runtime parameter \textit{eint\_switch}. If the internal energy is smaller than \textit{eint\_switch} times the kinetic energy, then the total energy is recomputed using the internal energy from Eq.~\ref{eq:inteng} and the velocities from the momentum equation. We have found that by setting  \textit{eint\_switch} $= 10^{-4}$, we are able to avoid any problems due to truncation error.

\begin{table}
\caption{Reactions in our non-equilibrium chemical model. \label{chem_model}}
\begin{tabular}{rlc}
\hline
No.\ & Reaction & Reference \\
\hline
1 & $\mH + \mH + {\rm grain} \rightarrow  \mHt + {\rm grain}$ & 1 \\
2 & $\mHt + \mH \rightarrow \mH + \mH + \mH$ & 2 \\
3 & $\mHt + \mHt \rightarrow \mH + \mH + \mHt$ & 3 \\
4 & $\mHt + \me \rightarrow \mH + \mH + \me$ & 4 \\
5 & $\mH + {\rm c.r.} \rightarrow \Hp + \me$ & See \S\ref{ics} \\ 
6 & $\mHt + {\rm c.r.} \rightarrow \mH + \mH$ & See \S\ref{ics} \\ 
7 & $\mHt  + {\rm c.r.} \rightarrow \mH + \Hp + \me$ & See \S\ref{ics} \\
8 & $\mH + \me \rightarrow \Hp + \me + \me$ & 5 \\  
9 & $\Hp + \me \rightarrow \mH + \gamma$ &  6 \\ 
10 & $\Hp + \me + {\rm grain} \rightarrow \mH + {\rm grain}$ & 7 \\ 
\hline
\end{tabular}
\medskip
\\
{\bf References}: 1: \citet{hm79}, 2: \citet{ms86}, 3: \citet{mkm98}, 
4: \citet{tt02}, 5:  \citet{a97}, 6: \citet{f92}, 7: \citet{wd01}
\end{table}

We treat the cooling coming from metals by assuming that the carbon, oxygen and silicon in the gas remain in the form of C$^{+}$, O and Si$^{+}$, respectively, as in the previous studies of \citet{gml07a,gml07b}. In practice, in the absence of photodissociating radiation (see below), we would expect carbon and silicon to rapidly recombine, and for the carbon to be converted to CO once the H$_{2}$ fraction becomes large. However, we know from previous work \citep{gc11a,gc11b} that the behaviour of the gas is not particularly sensitive to whether the dominant coolant is C$^{+}$ or CO. Cooling from C$^{+}$ alone can reduce the gas temperature to values around 15--20~K, and although CO cooling enables the gas to reach even lower temperatures (T $\sim$ 10 K), in realistic models of GMCs, the characteristic temperature of the fully molecular gas is generally in the range of 10 -- 20 K \citep{gc11b}. As the H$_{2}$ formation rate does not have a strong dependence on temperature, the approximate nature of our thermal treatment will have little influence on the H$_{2}$ formation rate in the gas. However, making this simplification allows us to minimize the computational requirements for our simulations by using a considerably simplified chemistry that follows only four species: free electrons, H$^{+}$, H, and H$_{2}$. We follow directly the fractional abundances of molecular hydrogen $x_{\rmn{H_{2}}}$ and ionized hydrogen $x_{\rmn{H^{+}}}$ (where these symbols denote the fraction of the available hydrogen found in these forms) by adding to the FLASH code an extra field variable for the mass density of each species. The abundances of the other two species - atomic hydrogen ($x_{\rmn{H}}$) and electrons ($x_{\rmn{e}}$) - are computed from the two conservation laws: conservation of charge
\begin{equation}
x_{\rmn{e}}=x_{\rmn{H^{+}}}+x_{\rmn{C^{+}}}+x_{\rmn{Si^{+}}}
\end{equation}
and conservation of the number of hydrogen nuclei
\begin{equation}
x_{\rmn{H}}=x_{\rmn{H,tot}}-x_{\rmn{H^{+}}}-x_{\rmn{H_{2}}}
\end{equation}
where $x_{\rmn{H,tot}}$ is the total abundance of hydrogen nuclei in all forms, and $x_{\rmn{C^{+}}}$ and $x_{\rmn{Si^{+}}}$ are the abundances of ionized carbon and silicon, respectively, which remain fixed throughout the simulations. These species undergo the reactions listed in Table~\ref{chem_model}. The radiative and chemical heating and cooling of the gas is modeled with a cooling function that contains contributions from the processes listed in Table~\ref{cool_model}.

\begin{table}
\caption{Processes included in our thermal model. \label{cool_model}}
\begin{tabular}{ll}
\hline
Process & Reference \\
\hline
C$^{+}$ fine structure cooling &  \citet{gml07a} \\
O fine structure cooling &  \citet{gf10} \\
Si$^{+}$ fine structure cooling &  \citet{gml07a} \\
$\mHt$ rovibrational lines & \citet{ga08} \\
Gas-grain energy transfer & \citet{hm89} \\
Recombination on grains & \citet{w03} \\
Atomic resonance lines & \citet{sd93} \\
$\mH$ collisional ionization& \citet{a97} \\
$\mHt$ collisional dissociation & See Table~\ref{chem_model} \\
$\mHt$ formation on dust grains & \citet{hm89} \\
Cosmic ray ionization & \citet{gl78}  \\  
\hline
\end{tabular}
\medskip
\\
\end{table}

We have also modified our treatment of the adiabatic index $\gamma$. \citet[]{b07} have recently pointed out that as the temperature of molecular gas increases, its specific heat capacity at constant volume, $c_{\upsilon}$, changes due to the fact that first the rotational and then the vibrational energy levels of H$_{2}$ become populated and that therefore $c_{\upsilon}$ cannot be considered constant and independent of temperature as has been often assumed in previous numerical studies of star formation. For this reason, we use a set of lookup tables constructed with the assumption that the H$_{2}$ ortho-to-para ratio has its thermal equilibrium value. In these tables, the specific internal energy $ \varepsilon$ is tabulated as a function of temperature $T$ and fractional abundance of H$_{2}$ ($x_{\rmn{H_{2}}}$), $T$ is tabulated as a function of $ \varepsilon$ and $x_{\rmn{H_{2}}}$, and the adiabatic index $\gamma$ is tabulated as a function of $ \varepsilon$ (or $T$) and $x_{\rmn{H_{2}}}$. To compute the required values for $\gamma$ or convert from $ \varepsilon$ to $T$ (or vice versa), we interpolate between the values stored in the tables.

To test our modified version of the FLASH code, we performed static and turbulent simulations using both our new FLASH implementation and our existing ZEUS-MP implementation \citep[]{gml07a, gml07b} of the same physics, and verified that the codes produced comparable results.

\subsubsection{Turbulent driving and hydrodynamics}

We have applied our chemistry model to simulations of forced supersonic turbulence driven by fully solenoidal (divergence-free or rotational) and fully compressive (curl-free or dilatational) forcing \citep[]{fed08b, fed09, fed10},  as two limiting cases to investigate the influence of the nature of the driving on the formation of H$_{2}$. These simulations use the piecewise parabolic method (PPM) \citep[]{ppm} implementation of the FLASH code to integrate the equations of hydrodynamics on 3D periodic uniform grids with 256$^{3}$ grid points. 

As a control parameter in our simulations, we use the rms velocity of the turbulence. We use this in preference to the rms Mach number because the latter quantity depends on the sound speed of the gas, and in our non-isothermal simulations this is not constant, but varies in both space and time. To excite a turbulent flow with a specified rms turbulent velocity, we include a forcing term $\mathbf{f}$ in the gas momentum equation
\begin{equation}
\frac{\partial {\mathbf v}}{\partial t} + (\mathbf{v} \cdot \nabla) \mathbf{v} = - \frac{\nabla P}{\rho} + \mathbf{f}.
\end{equation}
We model the random correlated stochastic forcing term $\mathbf{f}$ such that it varies smoothly in space and time using the Ornstein-Uhlenbeck (OU) process. The OU process is a well-defined stochastic process with a finite autocorrelation timescale $T$. It describes the evolution of the forcing term $\hat{f}$ in Fourier space ($k$-space) with the stochastic differential equation:
\begin{equation}
\rmn{d}\hat{f}(k,t)=f_{0}(k)\mathcal{\underline{P}}^{\zeta}(k)\rmn{d}W(t)-\hat{f}(k,t)\frac{\rmn{d}t}{T}
\label{eq:f}
\end{equation}
where $W(t)$ is a Wiener process, a random process that adds a Gaussian random increment to the vector field given in the previous time step $\rmn{d}t$, followed by the projection tensor $\mathcal{\underline{P}}^{\zeta}(k)$ in Fourier space. The projection operator reads
\begin{equation}
\mathcal{P}^{\zeta}_{ij}(k)=\zeta\mathcal{P}^{\perp}_{ij}(k)+(1-\zeta)\mathcal{P}^{\parallel}_{ij}(k)=\zeta\delta_{ij}+(1-2\zeta)\frac{k_{i}k_{j}}{|k|^2},
\end{equation}
where $\delta_{ij}$ is the Kronecker symbol, and $\mathcal{P}^{\perp}_{ij}=\delta_{ij}-k_{i}k_{j}/k^2$ and $\mathcal{P}^{\parallel}_{ij}=k_{i}k_{j}/k^2$ are the fully solenoidal and the fully compressive projection operators, respectively \citep[see e.g.][]{schm09,fed10}.

By changing the value of the parameter $\zeta$, we can determine the power of the compressive modes with respect to the total forcing power. For $\zeta$ = 1 in the projection operator, we obtain a purely solenoidal force field, and with $\zeta$ = 0, we obtain a purely compressive force field. Any combination of solenoidal and compressive modes can be constructed by choosing $\zeta\in [0, 1]$. 

The large-scale stochastic forcing that we use, as the one closest to the observational data \citep[]{oml02, br09}, models the kinetic energy input from large-scale turbulent fluctuations, breaking up into smaller structures. We thus drive the modes $k=[1,3]$ in units of $\frac{2\pi}{L}$, where $L$ is the box size. The forcing amplitude $A(k)$ has a parabolic dependence on $k$, such that most power is injected at $|\vec{k}|=2$ and $A(1)=A(3)=0$.

\subsection{Initial conditions}
\label{ics}

Using the forcing module described above, and starting from zero velocities, we excite turbulent motions in a box with 256$^3$ grid points and of side length $L$ = 20 pc, filled with initially uniform atomic gas, using periodic boundary conditions. We perform purely hydrodynamical simulations, and neglect any complications introduced by magnetic fields or the effects of self-gravity. The abundances for carbon, oxygen and silicon were taken from \citet[]{s00} and are: $x_{\rmn{C^{+}}} = 1.41 \times 10^{-4}$, $x_{\rmn{O}} = 3.16 \times 10^{-4}$ and $x_{\rmn{Si^{+}}} = 1.5 \times 10^{-5}$. We assume that the dust-to-gas ratio has the standard solar value, and fix the dust temperature at 10~K in every run. We adopt a rate $\zeta_{\rm H} = 10^{-17} \: {\rm s^{-1}}$ for the cosmic ray ionization of atomic hydrogen (reaction 5 in Table~\ref{chem_model}). In the case of molecular hydrogen, we assume that all of the H$_{2}^{+}$ ions produced in the reaction
\begin{equation}
\mHt + {\rm c.r.} \rightarrow {\rm H_{2}^{+}} + e^{-}
\end{equation}
are destroyed by dissociative recombination, yielding two hydrogen atoms, and so adopt a rate $\zeta_{\rm H_{2}, 6} = 2.22 \zeta_{\rm H}$ for reaction 6 that includes this contribution as well as that coming from direct dissociation of the H$_{2}$. For reaction 7, we adopt the rate $\zeta_{\rm H_{2}, 7} = 0.037 \zeta_{\rm H}$. In both cases, we assume that the ratio between the H$_{2}$ destruction rates and the ionization rate of atomic hydrogen is the same as given in \citet{umist}.

We perform two sets of simulations with different initial number densities: $n_{0}$ = 30 cm$^{-3}$ and $n_{0}$ = 300 cm$^{-3}$. For each initial density, we perform simulations with rms turbulent velocities of 0.4 km s$^{-1}$, 2 km s$^{-1}$ or 4 km s$^{-1}$, and examine both purely solenoidal and purely compressive forcing in each case, meaning that we perform a total of twelve simulations. We evolve each simulation for ten dynamical times $T=L/2v_{\rm rms}$. For the first two dynamical times, the chemistry module is switched off, and the turbulence is allowed to reach a statistically steady state \citep{fed09, fed10, pf10}. After that, we consider the chemical evolution and follow the gas for a further eight dynamical times. Note also that in our later discussion of the time evolution of the H$_{2}$ fraction, we take the time at which we switch on the chemistry module to be $t = 0$, meaning that the simulations run from $t = -2T$ until $t = 8T$.

For simplicity, we set the ambient radiation field strength to zero in all of our simulations, thereby avoiding the necessity of modeling the penetration of Lyman-Werner band photons into the simulation volume, and allowing us to focus purely on the influence of the turbulent density enhancements on the overall H$_{2}$ formation rate. We note that the mean column density through our low $n_{0}$ simulations is approximately $20 \: {\rm M_{\odot}} \: {\rm pc^{-2}}$, which is more than sufficient to adequately shield the H$_{2}$ in the gas against photodissociation \citep{kmt09}, provided that the incident radiation field is close to the standard Galactic value. We have shown in other work  \citep{gml10} that H$_{2}$ formation in clouds with surface densities of this value or higher is primarily limited by the time required to form the H$_{2}$, rather than by the influence of UV photodissociation. We therefore would not expect this omission to have a large impact on our results. At late times, we will tend to under-predict the amount of atomic hydrogen in the gas, and to over-predict the amount of H$_{2}$, particularly in our low density runs, but previous work suggests that the effect will be small \citep{gml10}. We note, however, that this approximation will break down for clouds immersed in UV radiation fields that are significantly stronger than the standard Galactic value (Glover, in preparation).
 
\subsection{Numerical resolution}

\begin{figure}
\centering
\includegraphics[width=7cm]{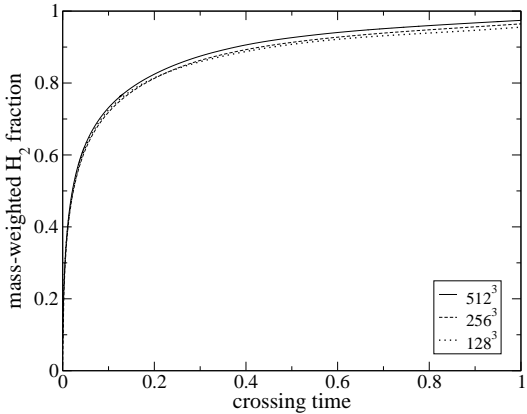}
\caption{Time evolution of the mass-weighted H$_2$ abundance in units of the turbulent crossing time for $n_{0}$ = 300 cm$^{-3}$ and $v_{\rm rms}$ = 4 km s$^{-1}$ in simulations with numerical resolutions of 128$^3$ grid points (dotted), 256$^3$ grid points (dashed) and 512$^3$ grid points (solid).}
\label{fig:H2res}
\end{figure}

\citet{gml07b} and \citet{mg10} examined the sensitivity of the H$_{2}$ formation timescale in simulations of the turbulent ISM to the numerical resolution of the simulation, using numerical resolutions ranging from 64$^3$ to 512$^3$ zones. They found that there was some dependence on the numerical resolution of the simulation at early times, owing to the ability of the higher resolution to better model the details of the highest density structures formed by the turbulence \citep[see][]{fed10,pf10}, although it should be noted that in these simulations the turbulence was not driven to a statistical steady-state before the switch-on of the chemistry, which will tend to exacerbate any resolution dependence. These previous studies found that although there remain some signs of resolution-dependence at 256$^3$ zones, the difference between the 128$^3$, 256$^3$ and 512$^3$ results is very small. However, these resolution tests were performed only for the case of solenoidal turbulence. Therefore, to test the sensitivity of H$_2$ formation to numerical resolution in the simulations with compressively driven turbulence, we have performed a resolution study for the run with $v_{\rm rms}$ = 4 km s$^{-1}$ and $n_{0}$ = 300 ${\rm cm^{-3}}$. This is the run in which the highest densities are produced, and so if this is well-resolved, then it is reasonable to assume that our lower density and lower $v_{\rm rms}$ runs will also be well-resolved. In our resolution study, we performed simulations with resolutions of $128^3$, $256^3$, and $512^3$ grid cells.

In Figure~\ref{fig:H2res}, we show how the mass-weighted mean abundance of H$_2$ (defined in section 3.1 below) evolves in runs with different resolution during the first crossing time. We see that there is almost no difference in the evolution of the H$_{2}$ abundance in the three simulations, and conclude that a numerical resolution of 256$^3$ grid cells should be enough to accurately model the growth of the H$_{2}$ fraction in our simulations.

\section{Results}

\subsection{Time dependence of H$_2$ abundance}

\begin{figure}
\centering
\includegraphics[width=7cm]{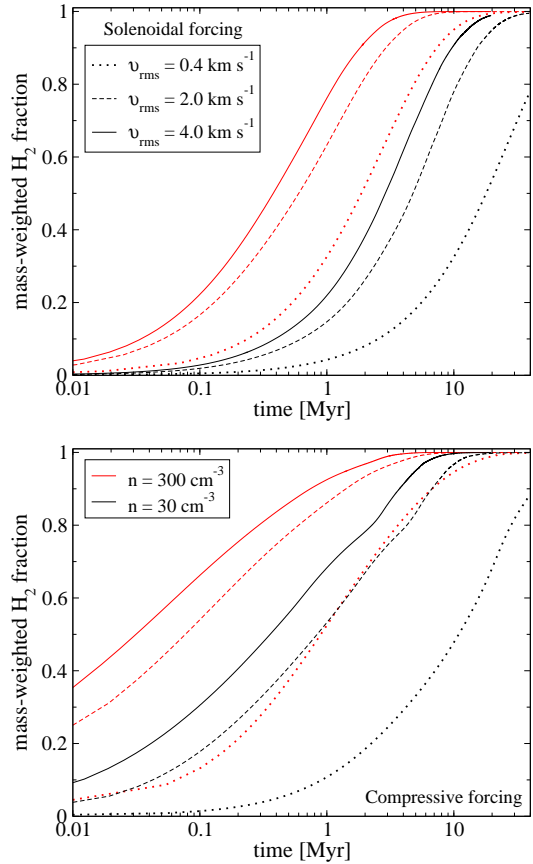}
\caption{Evolution with time of the mass-weighted mean H$_{2}$ fraction $\mass{x_{\rm H_{2}}}$ in runs with mean densities of 30~cm$^{-3}$ (black) and 300~cm$^{-3}$ (red). Three different values of the rms turbulent velocity $\upsilon_{\rmn{rms}}$ are considered: 0.4 km s$^{-1}$ (dotted), 2 km s$^{-1}$ (dashed) and 4 km s$^{-1}$ (solid). The upper panel shows the results for purely solenoidal forcing, while the lower panel shows the results for purely compressive forcing.}
\label{fig:H2time}
\end{figure}

To quantify the rate at which H$_2$ forms in our simulation we compute the mass-weighted mean molecular fraction, $\mass{x_{\rm H_{2}}}$, given by
\begin{equation}
\mass{x_{\rm H_{2}}}=\frac{\sum_{i,j,k}\rho_{\rm H_2}(i,j,k)\Delta V(i,j,k)}{M_{\rm H}}
\end{equation}
where we sum over all grid cells, and where $\rho_{\rm H_2}(i,j,k)$ is the mass density of H$_2$ in computational cell $(i,j,k)$, $\Delta V(i,j,k)$ is the volume of the cell $(i,j,k)$, $M_{\rm H}$ is the total mass of hydrogen present in the simulation. In Figure~\ref{fig:H2time}, we plot the evolution of $\mass{x_{\rm H_{2}}}$ as a function of time for both sets of runs, comparing different mean densities, rms velocities and types of driving. In Table~\ref{H2_time}, we give the time in Myr required for the mass-weighted mean molecular fraction to reach 50\% ($t_{50\%}$) and 90\% ($t_{90\%}$). 

\begin{figure*}
\centering
\includegraphics[scale=0.8]{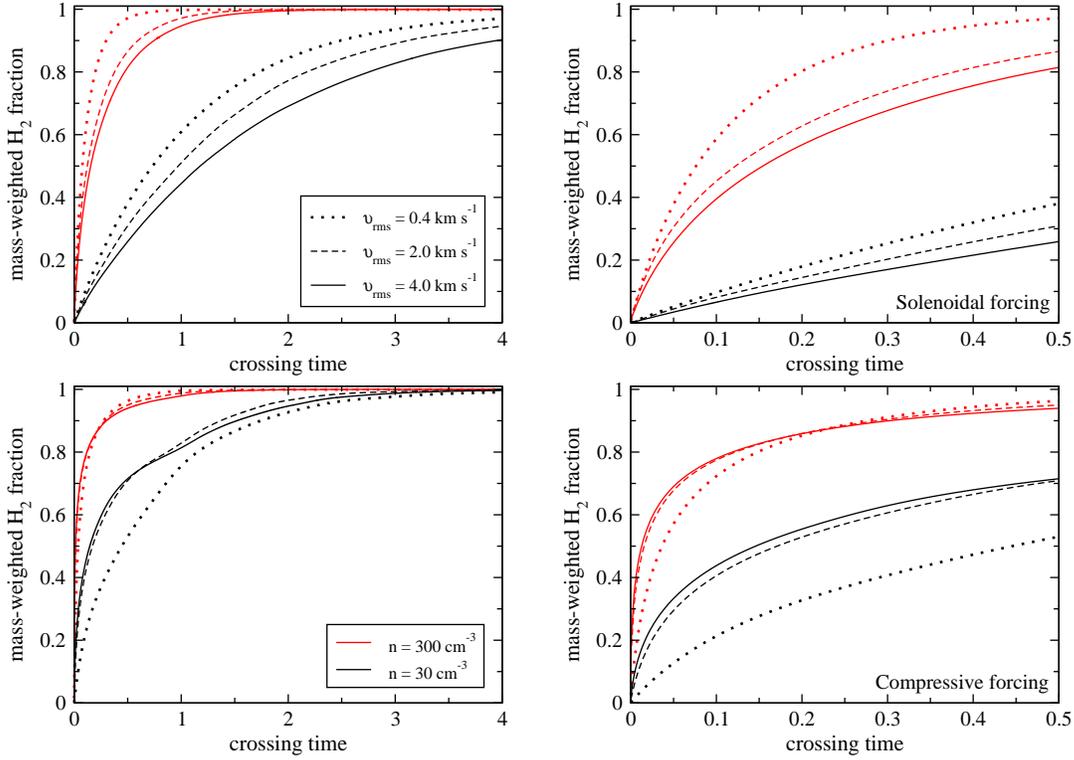}
\caption{As Figure~\ref{fig:H2time}, but showing the evolution of $\mass{x_{\rm H_{2}}}$ as a function of the turbulent crossing time $T$, rather than the absolute time. The left-hand panels show the evolution of $\mass{x_{\rm H_{2}}}$ from $t = 0$ to $t = 4T$, while the right-hand panels zoom in on the period between  $t = 0$ and $t = 0.5T$. As before, we plot results for three different values of the rms turbulent velocity -- 0.4 km s$^{-1}$ (dotted), 2 km s$^{-1}$ (dashed) and 4 km s$^{-1}$ (solid) -- and two different mean densities -- 30~cm$^{-3}$ (black) and 300~cm$^{-3}$ (red).}
\label{fig:H2cross}
\end{figure*}

Looking at the evolution of H$_2$ fraction with time in Figure~\ref{fig:H2time}, we see that the time required to convert a large fraction of the initial atomic hydrogen to molecular hydrogen decreases as we increase the density or the strength of turbulent driving, in line with the previous findings of \citet{gml07b}. Comparing the two panels, we see that compressively-driven turbulence leads to more rapid formation of H$_{2}$ than turbulence driven by solenoidal forcing. The difference is particularly pronounced at early times, and in runs with high rms velocities: for instance, $t_{50\%}$ is roughly a factor of ten smaller in the compressive run with $\upsilon_{\rmn{rms}}$=4 km s$^{-1}$ and $n_{0}$ = 300 cm$^{-3}$ than in the corresponding solenoidal run. At later times, the differences between the compressive and solenoidal runs become much smaller, with $t_{90\%}$ varying by less than a factor of three even in the most turbulent runs.

In Figure~\ref{fig:H2cross}, we show the evolution of the mass-weighted mean H$_2$ abundance as a function of the turbulent crossing time. Here we see that most of the dependence on the rms velocities vanishes when the time is measured in units of the crossing time. Regardless of the strength of the turbulence or the nature of the forcing, the molecular fraction reaches 50\% within only 0.1 -- 0.2 crossing times in the high density model. For the low density case it takes approximately 0.5 -- 1.0 crossing times to form the same amount of molecular gas, regardless of $\upsilon_{\rmn{rms}}$. 

Larger rms velocities yield more dense gas, resulting in a broader density PDF. On the other hand, they also lead to shorter turbulent crossing times, leaving less time for H$_2$ to form. As shown in Figure~\ref{fig:H2cross}, these two effects largely compensate for each other. In the solenoidal case, the latter effect dominates, and the H$_2$ formation timescale, in units of the crossing time, decreases with decreasing $\upsilon_{\rmn{rms}}$. In runs with compressive forcing, on the other hand, the increased width of the density PDF with increasing $\upsilon_{\rmn{rms}}$ is the dominant effect.

\subsection{Density and temperature distributions}

\begin{table}
\caption{Time in Myr when the gas becomes 50\% and 90\% molecular in all our runs. \label{H2_time}}
\begin{tabular}{l | c c | c c}
\hline
Initial number density & \multicolumn{2}{c}{$n_{0}$ = 30 cm$^{-3}$} & \multicolumn{2}{c}{$n_{0}$ = 300 cm$^{-3}$} \\
\hline
Solenoidal forcing & $t_{50\%}$ & $t_{90\%}$ & $t_{50\%}$ & $t_{90\%}$ \\
\hline
$\upsilon_{\rmn{rms}}$ = 0.4 km s$^{-1}$ & 17.94 & 60.97 & 1.91 & 7.36 \\
$\upsilon_{\rmn{rms}}$ = 2.0 km s$^{-1}$ & 4.79 & 15.30 & 0.64 & 2.96 \\
$\upsilon_{\rmn{rms}}$ = 4.0 km s$^{-1}$ & 2.88 & 9.67 & 0.38 & 1.83 \\
\hline
Compresive forcing & $t_{50\%}$ & $t_{90\%}$ & $t_{50\%}$ & $t_{90\%}$ \\
\hline
$\upsilon_{\rmn{rms}}$ = 0.4 km s$^{-1}$ & 10.95 & 42.73 & 0.9 & 6.74 \\
$\upsilon_{\rmn{rms}}$ = 2.0 km s$^{-1}$ & 0.87 & 6.74 & 0.11 & 1.44 \\
$\upsilon_{\rmn{rms}}$ = 4.0 km s$^{-1}$ & 0.36 & 3.73 & 0.036 & 0.74 \\
\hline
\end{tabular}
\medskip
\\
\end{table}

\begin{figure}
\centering
\includegraphics[width=7.3cm]{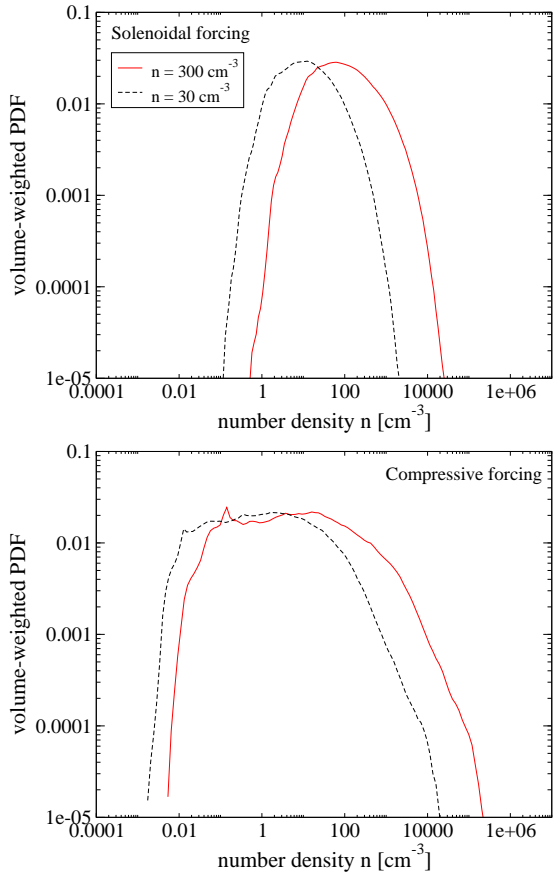}
\caption{Volume-weighted density PDF for solenoidal (top) and compressive (bottom) forcing at time $t=0.5$ crossing time in runs with $\upsilon_{\rmn{rms}}=2$ km s$^{-1}$. The red solid line presents the PDF in the run with mean density 300 cm$^{-3}$, while the black dashed line shows the PDF in the run with mean density 30 cm$^{-3}$.}
\label{fig:denspdf}
\end{figure}

\begin{figure}
\centering
\includegraphics [width=7cm]{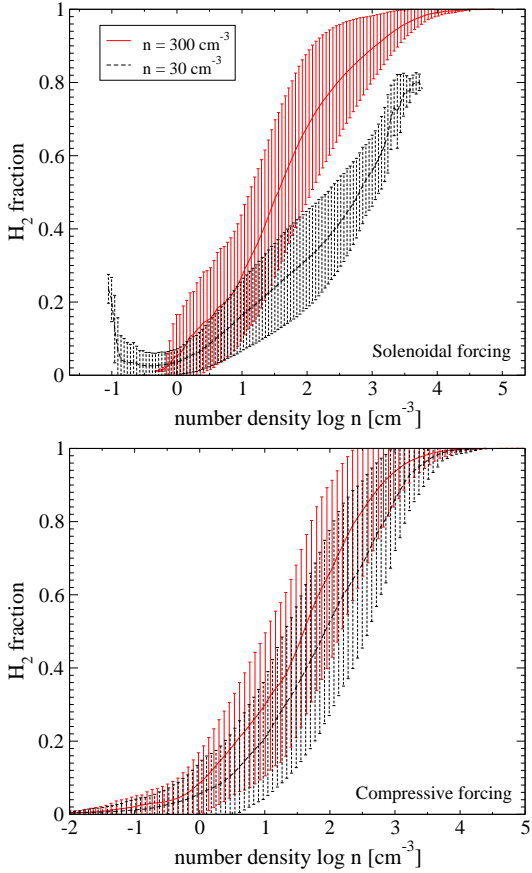}
\caption{Mean H$_2$ fraction, plotted as a function of the number density $n$ of the gas at time $t=0.5$ crossing time in runs with $\upsilon_{\rmn{rms}}=2$ km s$^{-1}$ that use solenoidal (top) and compressive (bottom) forcing. The red solid line indicates the runs with mean density $n_{0} = 300$~cm$^{-3}$, and the black dashed line indicates the runs with mean density $n_{0} = 30$~cm$^{-3}$. To compute these values, we binned the data by number density and computed the mean value of $x_{\rm H_{2}}$ for each bin. The standard deviation in the value of $x_{\rm H_{2}}$ in each bin is indicated by the error bars.}
\label{fig:H2hist}
\end{figure}

As Table~\ref{H2_time} demonstrates, the H$_2$ formation time does not scale linearly with changes in the density of the gas. We find that an increase in density by a factor of ten causes the gas to become 90\% molecular only 5 -- 8 times faster in the solenoidal case and 4 -- 6 times faster in the compressive case for the same rms turbulent velocities. The reason we see less dependence than one might naively expect is clear if we look at how the density distribution varies as we change the mean density $n_0$. In Figure~\ref{fig:denspdf} we plot a volume-weighted number density PDF at $t=0.5$ crossing times. As we decrease the density, the entire PDF moves to low densities. Most of the H$_2$ forms in dense gas, and so it is not surprising that reducing the amount of dense gas available has a significant effect on $x_{\rm H_{2}}$. However, the densest gas quickly becomes fully molecular and thereafter does not contribute to the total H$_2$ formation rate (see Fig.~\ref{fig:H2hist}), reducing the effect of density increase on the amount of formed H$_2$. We therefore find a smaller difference between the H$_2$ formation rates in the solenoidal and compressive runs than one might expect given the significant difference in the density PDF. 

In order to more quantitatively describe the H$_2$ distribution, we examine how the H$_2$ fraction varies with density. We compute $x_{\rm H_{2}}$ and $n$ for each of the cells in the simulation volume and then bin the data by number density. We then compute the mean and standard deviation for $x_{\rm H_{2}}$ in each bin. The resulting values at $t=0.5$ turbulent crossing times after the chemistry module is turned on are plotted in Figure~\ref{fig:H2hist}. We clearly see a considerable scatter in the value of $x_{\rm H_{2}}$ at a given density. However, there is still an obvious underlying trend in the distribution of  $x_{\rm H_{2}}$ with $n$, telling us that high density gas is more highly molecular, as expected \citep[e.g.][]{hws71}. At this point in the high density simulation the gas is almost fully molecular, whereas in the low density case $x_{\rm H_{2}}\simeq0.3$ for solenoidal and $x_{\rm H_{2}}\simeq0.7$ for compressive forcing (see Fig.~\ref{fig:H2cross}). Despite this, however, there are regions where the H$_2$ fraction is already much higher, and we can see that gas with a number density $n>10^3$ cm$^{-3}$ is already almost entirely molecular in all of the simulations.
 
\begin{figure}
\centering
\includegraphics [width=7cm]{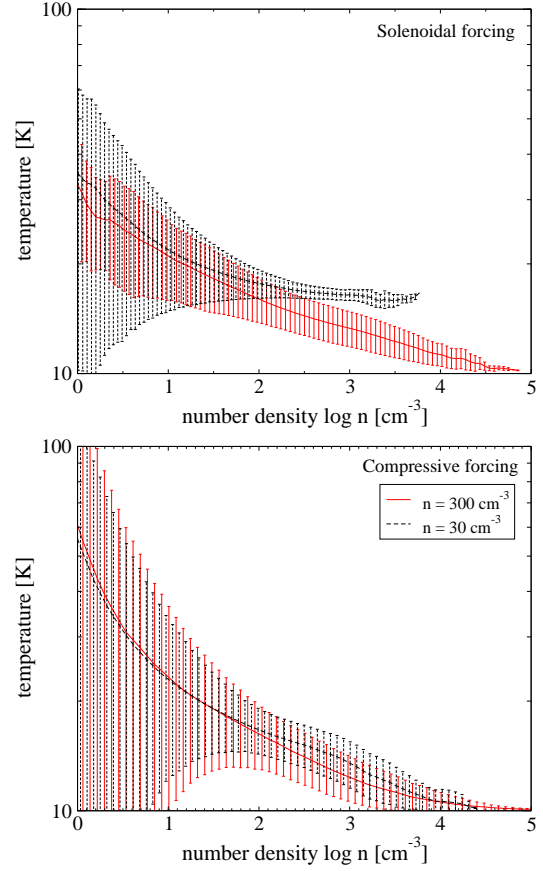}
\caption{Mean gas temperature plotted as a function of the number density $n$ at time $t=0.5$ crossing time in runs with $\upsilon_{\rmn{rms}}=2$ km s$^{-1}$ using solenoidal (top) and compressive (bottom) forcing. The red solid line indicates the run with mean density of 300 cm$^{-3}$ and the black dashed line indicates the run with mean density of 30 cm$^{-3}$. The data were binned in a similar fashion as for Figure~\ref{fig:H2hist}. The standard deviation in the mean value in each bin is also indicated.}
\label{fig:temphist}
\end{figure}

We also examine how the gas temperature varies as a function of number density in our simulations. Just as with the H$_2$ fraction above, we use the temperature output from our runs, bin it by number density $n$, and then compute the mean temperature and the standard deviation in the mean for each bin. We plot the resulting values again at $t=0.5$ turbulent crossing time in Figure~\ref{fig:temphist}. Strong shocks present in the turbulent simulations lead to high post-shock temperatures that can reach several thousand Kelvin. In low density gas, these shocks cause a significant scatter in the temperatures. In high density gas, their effect is less pronounced, owing to the significantly shorter cooling time. In the case of compressive forcing, the gas is found to have a wider range of densities than the gas in the case of solenoidal forcing. As discussed before, this is a result of the stronger compressions produced by the turbulent forcing. 

A final notable feature in the temperature distributions is the fact that in the low density solenoidal run, the temperature of the gas at $\log n\ge3.5$ is clearly higher than in the other runs. This occurs because in this run, there is still a significant quantity of atomic hydrogen present at these densities (see Fig.~\ref{fig:H2hist}), allowing heating due to H$_{2}$ formation to contribute significantly to the thermal balance of the gas. In the other runs, the atomic hydrogen fraction at these densities is very much smaller, and H$_{2}$ formation heating does not play a significant role in determining the gas temperature.

\subsection{Dependence on the density clumping factor}

As we are using periodic boundary conditions in our simulations, which prevent any of the H$_{2}$ molecules that form from escaping from the simulation volume, it is relatively straightforward to show that the evolution of the mass-weighted mean H$_{2}$ abundance with time is described by the following equation
\begin{equation}
\frac{{\rm d}\mass{x_{\rm H_{2}}}}{{\rm d}t} =  \mass{2R_{\rm H_{2}}(T, T_{\rm d}) x_{\rm H} n - D_{\rm H_{2}} x_{\rm H_{2}} n},
\end{equation}
where $R_{\rm H_{2}}(T, T_{\rm d})$ is the rate coefficient for H$_{2}$ formation on dust grains (reaction 1), and $D_{\rm H_{2}}$ is a destruction term depending on both temperature and density that accounts for the loss of H$_{2}$ in reactions 3, 4, 6 and 7 in Table~\ref{chem_model}. In practice, the impact of this destruction term is very small, unless $x_{\rm H} \ll x_{\rm H_{2}}$, and so to a good approximation
\begin{equation}
\frac{{\rm d}\mass{x_{\rm H_{2}}}}{{\rm d}t} \simeq  \mass{2R_{\rm H_{2}}(T, T_{\rm d}) x_{\rm H} n}.
\label{eq:h2}
\end{equation}
As it stands, Eq.~\ref{eq:h2} is not particularly useful, as in order to solve for the time dependence of $\mass{x_{\rm H_{2}}}$, we need to know how $R_{\rm H_{2}}$, $x_{\rm H}$, and $n$ are correlated, and how this correlation evolves with time. However, we can convert Equation~\ref{eq:h2} to a more useful form if we make a few further approximations. First, when the fractional ionization of the gas is small, as it is throughout our simulations, we have $x_{\rm H} \simeq 1 - x_{\rm H_{2}}$, and hence
\begin{equation}
\frac{{\rm d}\mass{x_{\rm H_{2}}}}{{\rm d}t} \simeq  \mass{2R_{\rm H_{2}}(T, T_{\rm d}) (1 - x_{\rm H_{2}}) n}.
\label{eq:h2b}
\end{equation}
Second, in our simulations we keep the dust temperature fixed, and we know that most of the gas has
a temperature that lies within the fairly narrow range of 10 -- 40 K (see Fig.~\ref{fig:temphist}). As
the dependence of $R_{\rm H_{2}}(T, T_{\rm d})$ on $T$ is weak when the temperature is low, we do not introduce a large error by treating the gas temperature (and hence $R_{\rm H_{2}}$) as if it were uncorrelated with the density, allowing us to write Equation~\ref{eq:h2b} as 
\begin{equation}
\frac{{\rm d}\mass{x_{\rm H_{2}}}}{{\rm d}t} \simeq  2R_{\rm H_{2}}(\mass{T}, T_{\rm d}) \mass{(1 - x_{\rm H_{2}}) n},
\label{eq:h2c}
\end{equation}
where $\mass{T}$ is the mass-weighted mean temperature.

\begin{figure*}
\centering
\includegraphics [width=14.5cm]{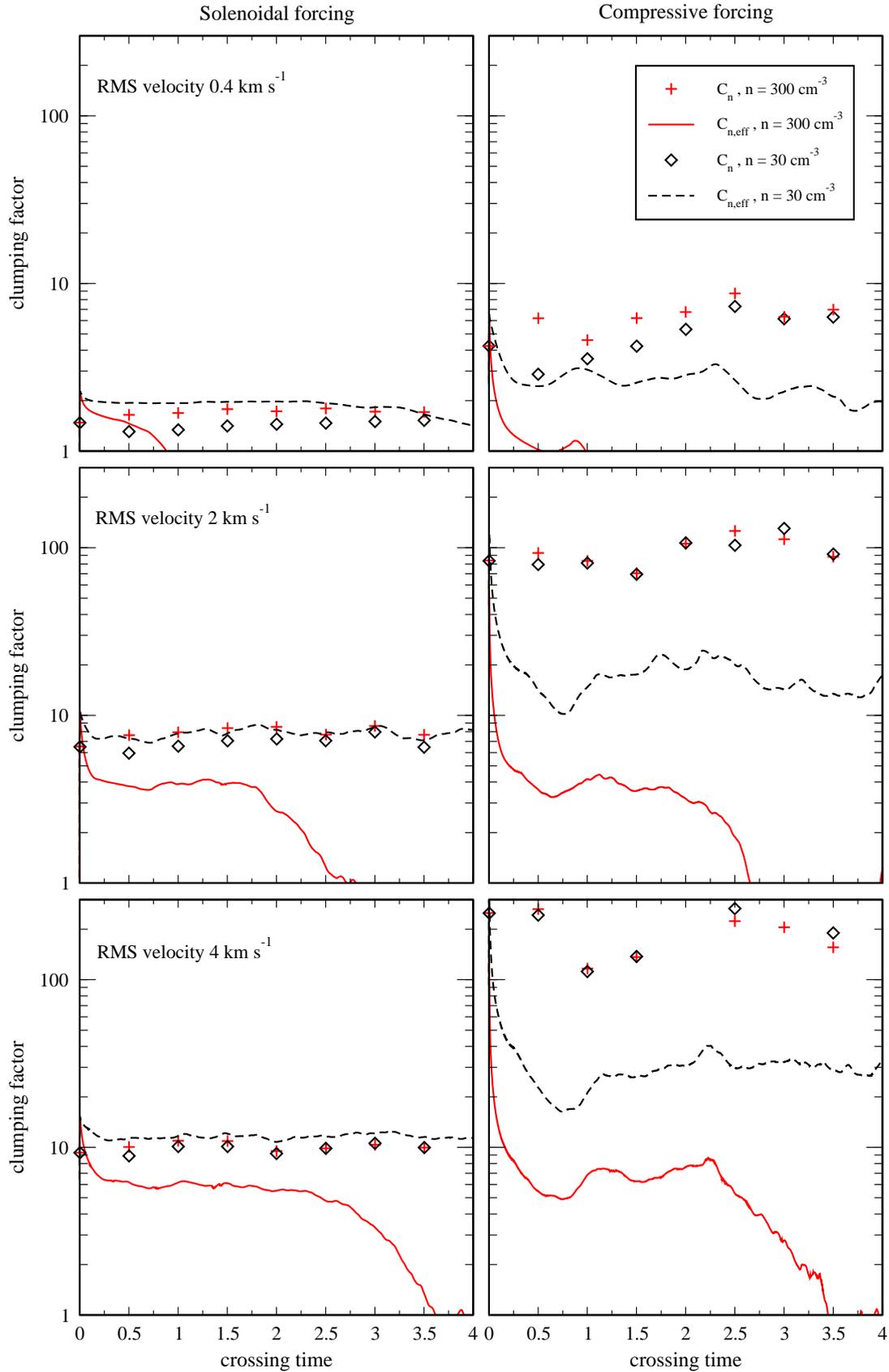}
\caption{Evolution of the effective clumping factor $C_{n, {\rm eff}}$ (lines) and the true clumping factor  $C_{n}$ (symbols) as a function of the turbulent crossing time $T$ in runs with mean densities of 30~cm$^{-3}$ (black) and 300~cm$^{-3}$ (red). We plot results for three different values of the rms turbulent velocity: 0.4 km s$^{-1}$ (top), 2 km s$^{-1}$ (middle) and 4 km s$^{-1}$ (bottom). The left-hand panels show the results for purely solenoidal forcing, while the right-hand panels show the results for purely compressive forcing.}
\label{fig:clump}
\end{figure*}

To proceed further, it is necessary to make an additional assumption regarding the correlation between the H$_{2}$ fraction
and the density. Given the presence of the turbulence, it is appealing to assume that this turbulence perfectly mixes the gas on a timescale much shorter than the chemical timescale. If we make this assumption, then we can 
treat $x_{\rm H_{2}}$ as being uncorrelated with density, allowing us to rewrite Equation~\ref{eq:h2c} as
\begin{eqnarray}
\frac{{\rm d}\mass{x_{\rm H_{2}}}}{{\rm d}t} & = &  2R_{\rm H_{2}}(\mass{T}, T_{\rm d}) \mass{(1 - x_{\rm H_{2}})} \mass{n} \\
 & = & 2R_{\rm H_{2}}(\mass{T}, T_{\rm d}) (1 - \mass{x_{\rm H_{2}}}) C_{n} \vol{n}, 
\label{eq:H2cn}
\end{eqnarray}
where $\vol{n}$ is the volume-weighted mean of $n$, defined as
\begin{equation}
\vol{n} \equiv \frac{1}{V} \int_{\rm V}  n  {\rm d}V.
\end{equation}
This quantity is related to the mass-weighted mean of $n$ by
\begin{eqnarray}
\mass{n} & = & \frac{1}{M} \int_{\rm V}  \rho n {\rm d}V, \\
 & = & \frac{1.4 m_{\rm H}}{M} \int_{\rm V} n^{2} {\rm d}V, \\
 & = & \frac{1.4 m_{\rm H}}{1.4 m_{\rm H} \vol{n} V} \vol{n^{2}}, \\
 & = & C_{n} \vol{n},
\end{eqnarray}
where $C_{n} \equiv \vol{n^{2}} / \vol{n}^{2}$ is the density clumping factor, and
where we have used the fact that $\rho = 1.4 m_{\rm H} n$, and hence that $M \equiv  \vol{\rho}
V = 1.4 m_{\rm H} \vol{n} V$. 

Equation~\ref{eq:H2cn} demonstrates that if our assumption of rapid mixing of the H$_{2}$ were true, then the evolution of the mass-weighted mean H$_{2}$ fraction in a gas cloud would be related in a very simple fashion to the mean density of the cloud and its density clumping factor. This fact has been used by \citet{gtk09} as the basis of a simple sub-grid scale model of H$_{2}$ formation for cosmological simulations, or for other large-scale simulations without sufficient resolution to model the small-scale structure within molecular clouds. They write the formation rate of H$_{2}$ in a similar form to Eq.~\ref{eq:H2cn}, and argue that $C_{n} \sim 3$--10 in typical turbulent clouds. \citet{gk11} further developed this idea, and showed that this sub-grid model does a good job of reproducing the dependence of the average atomic and molecular gas surface densities on the total hydrogen surface density that is observed in nearby spiral galaxies \citep{wb02}, and the dependence of the mean H$_{2}$ fraction on the total hydrogen column density observed in our own Galaxy \citep{gill06,wolf08}. 

However, the fact that we see a clear correlation between $x_{\rm H_{2}}$  and $n$ in our simulations (see Fig.~\ref{fig:H2hist}) implies that the assumption of rapid mixing that we used to derive Equation~\ref{eq:H2cn} is incorrect. In reality, it takes roughly one-third of a turbulent crossing time to fully mix material from overdense clumps into their lower density surroundings for solenoidal turbulence \citep[]{fed08a}, and potentially longer than this for compressive turbulence. Therefore, a prescription such as that in Eq.~\ref{eq:H2cn} will overestimate the H$_{2}$ formation rate.  

\begin{figure*}
\centering
\includegraphics [width=14.5cm]{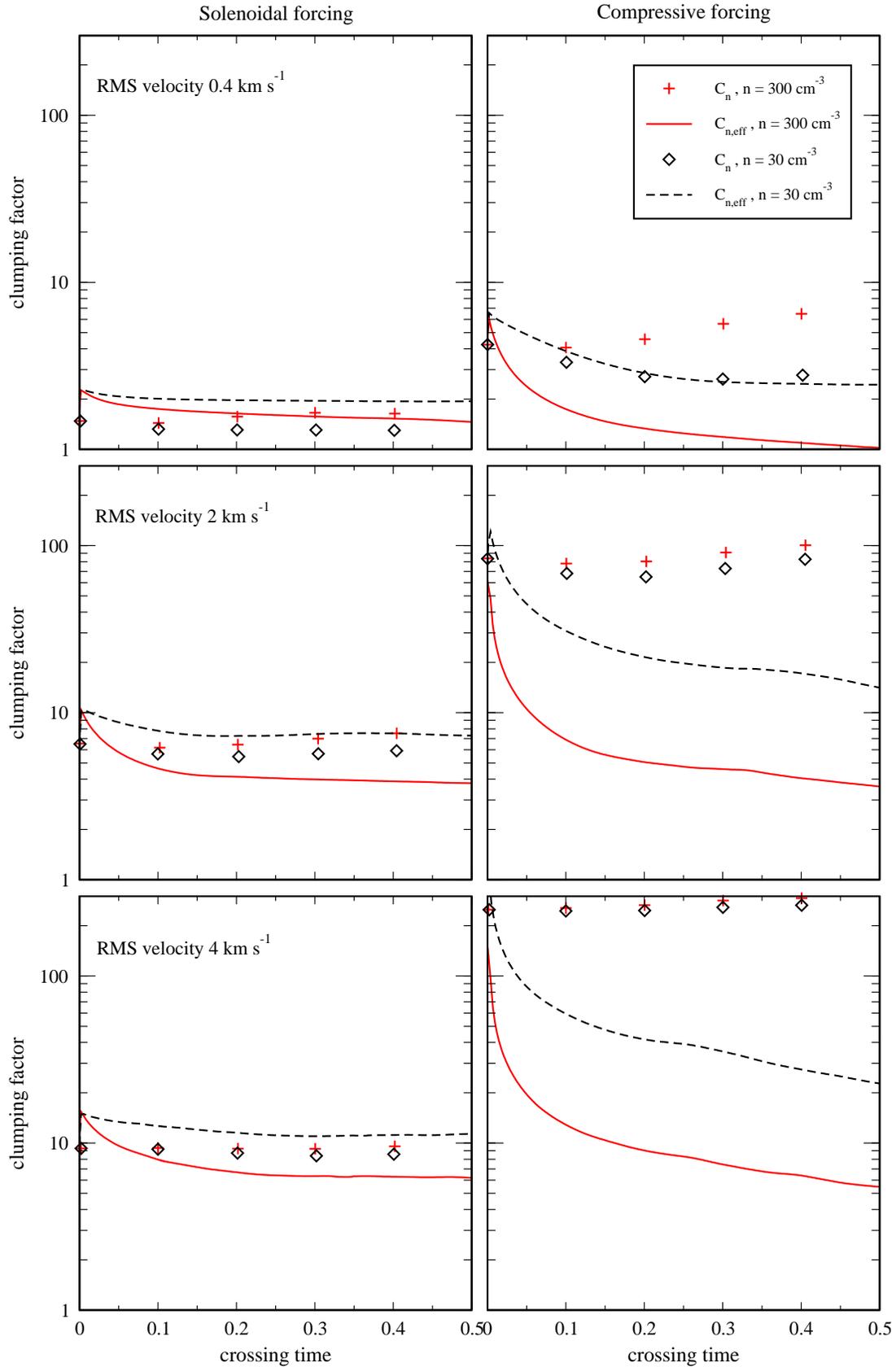}
\caption{As Figure~\ref{fig:clump}, but showing an expanded view of the first 0.5 crossing times. As before, three different values of the rms turbulent velocity $\upsilon_{\rmn{rms}}$ are considered: 0.4 km s$^{-1}$ (top), 2 km s$^{-1}$ (middle) and 4 km s$^{-1}$ (bottom) - and two different mean densities - 30~cm$^{-3}$ (black) and 300~cm$^{-3}$ (red). The left-hand panels show the results for purely solenoidal forcing, while the right-hand panels show the results for purely compressive forcing. }
\label{fig:clumpzoom}
\end{figure*}

Our present simulations of  solenoidal and compressive turbulence provide a useful test-bed for quantifying the extent to which Eq.~\ref{eq:H2cn}, and by extension the Gnedin et~al.\ sub-grid model, overestimates the H$_{2}$ formation rate. To do this, we define an `effective' density clumping factor
\begin{equation}
C_{n, {\rm eff}} = \frac{{\rm d}\mass{x_{\rm H_{2}}} / {\rm d}t}{2R_{\rm H_{2}}(\mass{T}, T_{\rm d}) (1 - \mass{x_{\rm H_{2}}}) \vol{n}},
\end{equation}
and compute how it evolves with time in each of our simulations, using our results for 
$\mass{x_{\rm H_{2}}}$ and $\mass{T}$ discussed earlier. We then compare this with the true density
clumping factor $C_{n}$ computed at a number of different times during the simulations. The results of
this comparison are plotted in Figure~\ref{fig:clump} (which shows the evolution between 0 and 4 crossing times) and Figure~\ref{fig:clumpzoom} (which shows an expanded view of the first 0.5 crossing times).

We see that at the very earliest times in the runs, there is a reasonable level of agreement between our inferred effective clumping factor $C_{n, {\rm eff}}$ and the measured clumping factor $C_{n}$. Our computed values of $C_{n, {\rm eff}}$ are typically some 20--40\% larger than $C_{n}$, but an error of this magnitude is plausibly explained by our use of the mass-weighted mean temperature in our calculation of $R_{\rm H_{2}}$: in reality, the dense gas, whose contribution initially dominates the H$_{2}$ formation rate, will generally be colder than this mean temperature.

However, this initial level of agreement between $C_{n, {\rm eff}}$ and $C_{n}$ is very quickly lost in most of the runs. In all of the simulations, the true clumping factor $C_{n}$ remains approximately constant, varying by at most a factor of two in the compressive case, and by much less than this in the solenoidal case. On the other hand, in most of the runs, $C_{n, {\rm eff}}$ decreases rapidly with time; only in the low density solenoidal model it does remain approximately constant during the lifetime of the simulation. The strong and almost immediate decrease of the effective clumping factor visible in Figures~\ref{fig:clump} and \ref{fig:clumpzoom} is caused by the increase in the H$_{2}$ abundance in the dense gas. As the dense regions that initially dominate the H$_{2}$ formation rate become almost fully molecular, their contribution decreases rapidly, causing a significant fall in the mean H$_{2}$ formation rate within the simulation, and hence a significant decrease in $C_{n, {\rm eff}}$. This effect is particularly pronounced in the compressively-forced runs, owing to their broad density PDFs. If we closely compare the results plotted in Figure~\ref{fig:clumpzoom}  with the time evolution of the H$_2$ fraction shown in Figure~\ref{fig:H2cross}, we can see that the \citet[]{gtk09} approach starts to break down when the gas is about 30\% molecular. In the high density solenoidal runs,  the H$_2$ formation rate is almost immediately overestimated by a factor of 2, while in the compressive runs, the rate is overestimated by a factor of 4 in the low density case, and by a factor of 10 in the high density case. 

It is clear from this analysis that in most cases there is no simple way to relate the mean number density of the gas and the current mass-weighted mean H$_2$ abundance to the current H$_{2}$ formation rate, given the strong time variation that we see in $C_{n, {\rm eff}}$. This time variation is absent only when the characteristic H$_{2}$ formation timescale is longer than a turbulent crossing time, as is the case in our low-density solenoidal runs, as only in this case is our assumption of rapid turbulent mixing justified. One must therefore be careful when using the \citet{gtk09} sub-grid model to describe the H$_{2}$ formation rate in numerical simulations. 

\section{Summary}

We have presented the results of a study of H$_2$ formation in the turbulent ISM that examines the influence of the amplitude and mode of both solenoidal and compressive turbulent driving. We have performed high-resolution 3D hydrodynamic simulations using the massively parallel code FLASH, which we have modified to include a detailed treatment of atomic/molecular cooling and the most important hydrogen chemistry. Even though the chemical network we use is significantly simplified compared to the most detailed models available, it performs with acceptable accuracy for our purposes. We have performed simulations with numerical resolutions of 128$^{3}$, 256$^{3}$ and 512$^{3}$ zones, and have demonstrated that our results are well-converged in our 256$^{3}$ runs. Our results also serve as a proof-of-concept application for our implementation of our non-equilibrium chemical model within the FLASH adaptive mesh refinement code.

We find that with both compressively and solenoidally driven turbulence, molecular hydrogen forms faster in gas with a higher mean density, or an environment with stronger turbulence. Although initially (during the first million years), H$_2$ formation is significantly faster with compressive turbulence than with solenoidal turbulence, at later times the differences become smaller, with the time taken to reach a molecular hydrogen fraction of 90\% varying by at most a factor of three between the compressive and solenoidal runs. In almost all of our simulations, the gas becomes highly molecular within a much shorter time than the 10--20~Myr that would plausibly be required to assemble the cloud from the diffuse ISM \citep{bphv99,el00,hartmannetal01}.

We have also shown that when time is measured in the units of turbulent crossing time, the H$_2$ formation timescale becomes much less dependent on the strength of the turbulence. Increasing the strength of the turbulence produces more dense gas and reduces the time taken to form H$_{2}$. However, it also reduces the turbulent crossing time of the gas. In the solenoidal case, the reduction in the turbulent crossing time is the dominant effect, and so H$_{2}$ formation takes {\em longer} (in units of the crossing time) as we increase $\upsilon_{\rmn{rms}}$. On the other hand, in the compressive case, the broadening of the density PDF is the dominant effect, and increasing $\upsilon_{\rmn{rms}}$ leads to a moderate decrease in the H$_{2}$ formation timescale measured in units of the crossing time.

The differences we have found between the compressive and solenoidal runs can largely be understood by considering the differences in the density PDFs in Figure~\ref{fig:denspdf}. Compressive forcing produces a much wider spread of densities than solenoidal forcing, and since the H$_{2}$ formation rate per unit volume scales almost linearly with density when $x_{\rm H_{2}}$ is small, this allows the compressive runs to form H$_{2}$ much more rapidly at early times. However, rapid H$_2$ formation in the dense gas leads to its conversion to fully molecular form, at which point it no longer contributes to the total H$_{2}$ formation rate. This phenomenon occurs in both the solenoidal and the compressive runs, but has a greater effect in the compressive runs owing to the faster initial H$_{2}$ formation rate in these runs. 

Finally, we have also used the results of our study to show that the \citet{gtk09} prescription for correcting for the influence of  unresolved density fluctuations on the H$_{2}$ formation rate in large-scale Galactic or cosmological simulations must be used with caution. The \citet{gtk09} prescription assumes rapid gas mixing, when in reality it takes about one-third of a turbulent crossing time to mix the material from overdense clumps into the low density regions in the case of solenoidal forcing, and possibly even longer in the case of compressively-driven turbulence \citep[]{fed08a}. We have shown that the effective clumping factor calculated with the assumption of rapid mixing over-predicts the H$_2$ formation rate. In the case of high density and strong compressive forcing, the H$_{2}$ formation rate can be overestimated by more than an order of magnitude at all but the very earliest times. For applications where one simply wants to determine which regions of the ISM become H$_{2}$-dominated (i.e.\ more than 50\% molecular) and how quickly this occurs, their approach remains reasonably accurate, since $C_{n, {\rm eff}}$ shows little variation while $\mass{x_{\rm H_{2}}}$ remains small. On the other hand, if one is interested in the final, equilibrium state of the gas \citep[as in e.g.][]{kg10}, then this approach may be problematic, as it will systematically over-predict the H$_{2}$ formation rate in highly molecular regions, with the result that the H$_{2}$ abundance will reach equilibrium too rapidly.

\section*{Acknowledgments}

We thank M.-M. Mac Low, S. Walch, and M. Wolfire for interesting discussions and valuable comments on the present work. We thank the anonymous referee for insightful comments. M.M. acknowledges financial support by the International Max Planck Research School for Astronomy and Cosmic Physics at the University of Heidelberg (IMPRS-HD) and the Heidelberg Graduate School of Fundamental Physics (HGSFP). The HGSFP is funded by the Excellence Initiative of the German Research Foundation DFG GSC 129/ 1. C.F.~has received funding from the European Research Council under the European Community's Seventh Framework Programme (FP7/2007-2013 Grant Agreement no.~247060) for the research presented in this work. S.C.O.G,~C.F.,~and R.S.K.\ acknowledge financial support from the {\em  Baden-W{\"u}rttemberg Stiftung} via the program International Collaboration II (grant P-LS-SPII/18) and from the German {\em Bundesministerium f\"{u}r Bildung und Forschung} via the ASTRONET project STAR FORMAT (grant 05A09VHA). R.S.K. furthermore gives  thanks for subsidies from the {\em Deutsche Forschungsgemeinschaft}  under grants no.\ KL 1358/1, KL 1358/4, KL 1359/5, KL 1358/10, and KL 1358/11, as well as from a Frontier grant of Heidelberg University sponsored by the German Excellence Initiative. The simulations used computational resources from the HLRBII project pr32hu at Leibniz Rechenzentrum Garching. The software used in this work was in part developed by the DOE-supported ASC/Alliance Center for Astrophysical Thermonuclear Flashes at the University of Chicago.

\bsp

\label{lastpage}

\end{document}